\documentclass[twoside,11pt]{article}

\usepackage{jmlr2e}

\usepackage{blindtext}
\usepackage{xcolor}
\usepackage{tikz}
\usepackage{lmodern,adjustbox,booktabs,multirow}
\usepackage{pifont}
\usepackage{float}
\usepackage{tabularx}
\usepackage{bbm}
\usepackage{amsmath}
\usepackage[linesnumbered,ruled]{algorithm2e}

\newcommand{\netlib}{\href{https://www.netlib.org/lp/data/}{NetLib} }

\usepackage{lastpage}

\ShortHeadings{Benny Sun and Yuansi Chen}{Open Source MCMC Sampling}
\firstpageno{1}

\newcommand{\greencheck}{{\color{green}\ding{51}}}
\newcommand{\redcross}{{\color{red}\ding{55}}}
\newcommand{\package}{\textsf{PolytopeWalk}\xspace}
\newcommand{\cC}{\mathcal{C}}

\newcommand{\cF}{\mathcal{F}}

\newcommand{\defn}{:=}

\long\def\comment#1{}
\definecolor{battleshipgrey}{rgb}{0.52, 0.52, 0.51}
\definecolor{darkgray}{rgb}{0.66, 0.66, 0.66}
\definecolor{darkgreen}{rgb}{0.0, 0.2, 0.13}
\definecolor{darkspringgreen}{rgb}{0.09, 0.45, 0.27}
\definecolor{dukeblue}{rgb}{0.0, 0.0, 0.61}
\definecolor{olivedrab7}{rgb}{0.24, 0.2, 0.12}
\definecolor{darkblue}{rgb}{0.0, 0.0, 0.55}
\definecolor{darkscarlet}{rgb}{0.34, 0.01, 0.1}
\definecolor{candyapplered}{rgb}{1.0, 0.03, 0.0}
\definecolor{ao(english)}{rgb}{0.0, 0.5, 0.0}
\definecolor{applegreen}{rgb}{0.55, 0.71, 0.0}

\DeclareMathOperator{\diag}{diag}

\newcommand{\real}{\ensuremath{\mathbb{R}}}

\newcommand{\braces}[1]{\left\{ #1 \right \}}
\newcommand{\abss}[1]{\left| #1 \right |}

\newcommand{\tp}{^\top}

\newcommand{\bmat}[1]{\begin{bmatrix} #1 \end{bmatrix}}

\begin{document}
\setcitestyle{numbers, open={f}, close={f}}

\title{\package: Sparse MCMC Sampling over Polytopes}

\author{\name Benny Sun\email benny.sun@duke.edu \\
       \addr Department of Statistics\\
       Duke University\\
       Durham, NC 27708, USA
       \AND
       \name Yuansi Chen \email yuansi.chen@stat.math.ethz.ch \\
       \addr Department of Mathematics\\
       ETH Zurich\\
       Zurich 8092, Switzerland}
\editor{Joan Durso}
\maketitle

\begin{abstract}
High dimensional sampling is an important computational tool in statistics and other computational disciplines, with applications ranging from Bayesian statistical uncertainty quantification, metabolic modeling in systems biology to volume computation. We present $\textsf{PolytopeWalk}$, a new scalable Python library designed for uniform sampling over polytopes.  The library provides an end-to-end solution, which includes preprocessing algorithms such as facial reduction and initialization methods. Six state-of-the-art MCMC algorithms on polytopes are implemented, including the Dikin, Vaidya, and John Walk. Additionally, we introduce novel sparse constrained formulations of these algorithms,  enabling efficient sampling from sparse polytopes of the form $\mathcal{K}_2 = \{x \in \mathbb{R}^d \ | \  Ax = b, x \succeq_k 0\}$. This implementation maintains sparsity in $A$, ensuring scalability to high dimensional settings $(d > 10^5)$. We demonstrate the improved sampling efficiency and per-iteration cost on both \netlib datasets and structured polytopes. \package is  available at \textsf{github.com/ethz-randomwalk/polytopewalk} with documentation at \textsf{polytopewalk.readthedocs.io/}.
\end{abstract}


\section{Introduction}
High dimensional sampling is a fundamental problem in many computational disciplines such as statistics, probability, and operation research. For example, sampling is applied in portfolio optimization \cite{DBLP:journals/corr/abs-1803-05861}, metabolic networks in systems biology \cite{COBRA} and volume approximation over convex shapes from \cite{Simonovits2003}. Markov chain Monte Carlo (MCMC) sampling algorithms offer a natural and scalable solution to this problem. These algorithms construct a Markov chain whose stationary distribution matches the target distribution. By running the chain for a sufficient number of steps to ensure mixing, MCMC algorithms can efficiently generate approximately independent samples close to the target distribution, while not suffering from the curse of dimension issues.

In this work, we focus on the problem of sampling from a uniform distribution over a polytope. Let $A \in \mathbb{R}^{n \times d}$, $b \in \mathbb{R}^n$ and define  $x \succeq_k y$ to mean that the last $k$-coordinates of $x$ are greater than or equal to the corresponding coordinates of $y$, i.e., $\{x_{d-k+1} - y_{d-k+1} \ge 0, ... , x_{d} - y_{d} \ge 0\}$. Depending on whether we allow equality constraints, the sampling problem can be formalized in two forms:
\begin{enumerate}
    \item The full-dimensional form:
    \begin{align}
        \mathcal{K}_1 = \{x \in \mathbb{R}^d \ | Ax \le b\},
        \label{eq:full_dim}
    \end{align}
    where $\mathcal{K}_1$ is specified via $n$ inequality constraints. 
    \item The constrained form:
    \begin{align}
        \mathcal{K}_2 = \{x \in \mathbb{R}^d \ | \ Ax = b, x \succeq_k 0\},
        \label{eq:constrained}
    \end{align}
    where $\mathcal{K}_2$ is specified via $n$ equality constraints and $k$ coordinate-wise inequality constraints. 
\end{enumerate}
Large polytopes with sparse constraints are common in many applications. For instance, the largest human metabolic network RECON3D is modeled as a $13543$-dimensional sparse polytope \cite{10.1093/nar/gkv1049}. Additionally, many linear programming datasets from \netlib are naturally in the constrained form, where $A$ matrix is sparse. These applications motivate the need for MCMC algorithms that leverage the large and sparse $\mathcal{K}_2$ formulation. In a similar spirit of the recent package for running Riemannian Hamiltonian Monte Carlo algorithm on large sparse polytopes,   \cite{DBLP:journals/corr/abs-1911-05656}, in our paper, we develop novel interior-point-method-based MCMC algorithms optimized for large and sparse constrained polytopes. By exploiting sparsity, our algorithms scale well in both per-iteration cost and overall sampling efficiency as a function of increasing dimension, enabling effective sampling from polytopes with dimensions exceeding $10^5$.

Interior-point-method-based MCMC sampling algorithms on a polytope are modifications of the Ball Walk \cite{vempala2005}, incorporating key concepts from interior-point methods in optimization. These algorithms operate in two primary steps. First, the algorithm generates a proposal distribution whose covariance matrix is state-dependent and equal to the inverse of the Hessian matrix of a specified barrier function, capturing the local geometry of the polytope. Second, the algorithm employs the Metropolis-Hastings accept-reject step to ensure that its stationary distribution is uniform on the polytope. Using a state-dependent proposal distribution that adapts to the polytope's local geometry, these MCMC algorithms achieve an improved mixing rate.  Specific algorithms in this class include the Dikin Walk \cite{DBLP:journals/corr/SachdevaV15}, Vaidya Walk \cite{JMLR:v19:18-158}, John Walk \cite{JMLR:v19:18-158}, and Lee Sidford Walk \cite{DBLP:journals/corr/abs-1911-05656}, with theoretical guarantees therein.  Each of these methods leverages different barrier functions and are specialized for sampling distributioned truncated on a polytope.

In \package, we implement 4 interior-point-method-based MCMC sampling algorithms in both the sparse, constrained formulation and the full-dimensional form. \package is one of the few open-source packages that enables users to leverage the sparsity in the specified polytope to speed up calculations. We demonstrate our packages' comparative advantages against $\textsf{Volesti}$ \cite{Chalkis_2021} in sampling time by testing our packages on both real-life \netlib datasets and structured families of polytopes. Finally, we provide an an open-source implementation of the Facial Reduction algorithm, used to handle degeneracy in polytopes.

\section{Package Overview}
\package is an open-source library written in C++ with Python wrapper code. There are 3 main components of the package: preprocessing, sampling, and post processing. Our package comes with many promising features.

\begin{figure}[ht]
    \centering
    \includegraphics[width = 16cm] {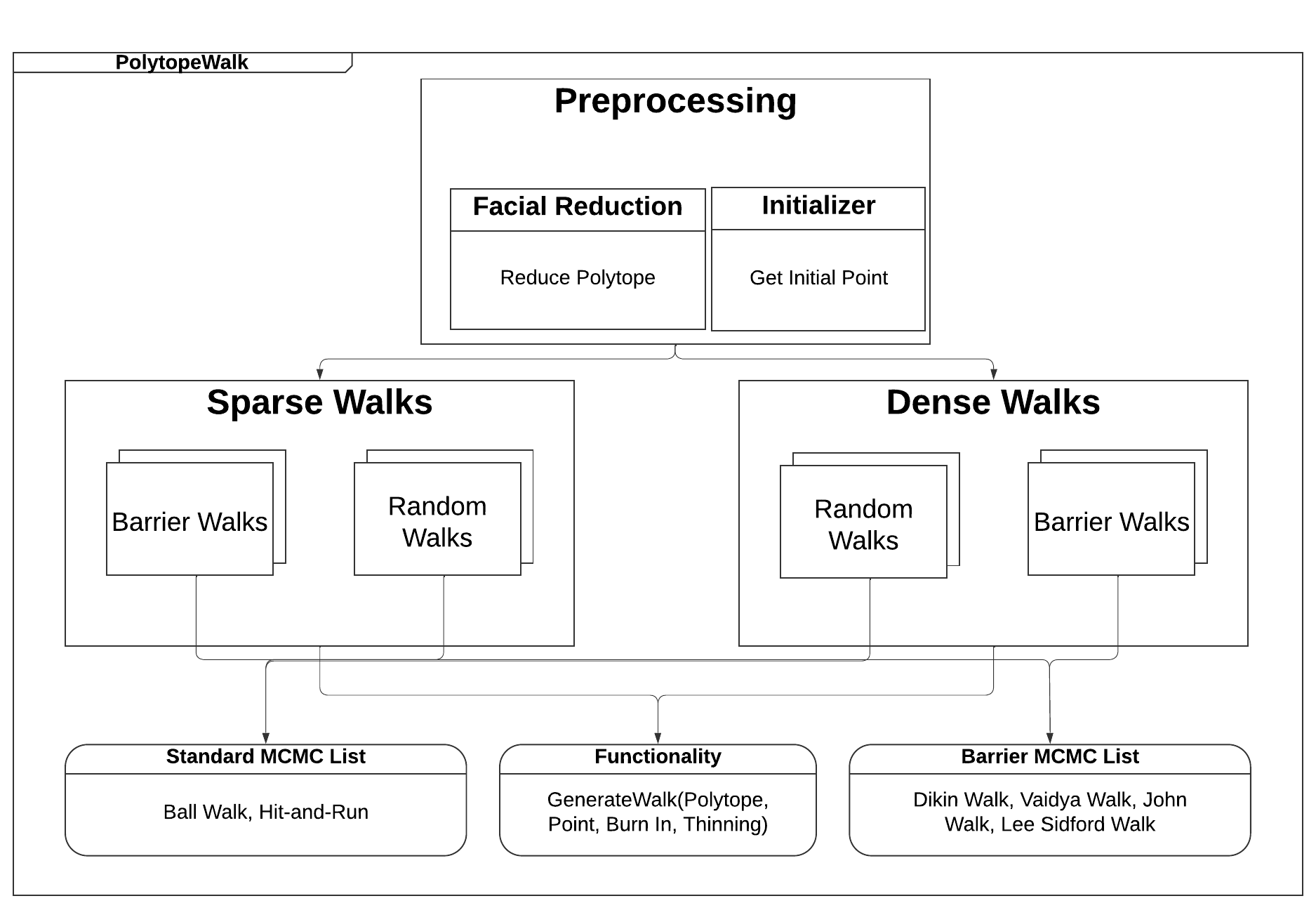}
    \caption{Code Structure of Package}
    \label{fig:enter-label}
\end{figure}
\paragraph{Comprehensive Initialization} We develop several algorithms to preprocess polytope inputs, simplifying the dimension space and providing an initial sampling point. The facial reduction algorithm removes degeneracy in the input space, returning a polytope in a smaller $\mathcal{K}_2$ form and preserving its sparsity. Therefore, the facial reduction algorithm, which has not been developed in open-source code, provides a stronger theoretical guarantee of always terminating in the aforementioned form, an improvement over less clear methods \cite{im2023revisiting}.

\paragraph{Fast Runtimes} \package provides accelerated MCMC sampling algorithms in both the $\mathcal{K}_1$ formulation~\eqref{eq:full_dim} and $\mathcal{K}_2$ formulation~\eqref{eq:constrained}. The source code is written in low-level C++ with $\href{https://eigen.tuxfamily.org/}{\textsf{Eigen}}$ for optimized linear algebra operations (both sparse and dense),  $\href{https://www.gnu.org/software/glpk}{\textsf{glpk}}$ for a fast sparse linear programming solver and $\href{https://github.com/pybind/pybind11}{\textsf{pybind}}$ to enable Python binding. In the experiments section, we demonstrate its superiority in runtimes over the \textsf{Volesti} package.

\paragraph{Documentation} \package includes documentation and testing for the entire library. Unit tests call every function in the library ensuring over 90\% code coverage. Documentation in the C++/Python source code describes the purpose of each function, inputs, and output. \package also includes an example section with tutorial-style notebooks in place. 

\subsection{Random Walk Algorithms}
There are 2 main submodules: $\textsf{dense}$ and $\textsf{sparse}$. The $\textsf{dense}$ module includes implementations of the 6 random walks algorithms in the full-dimensional formulation. The $\textsf{sparse}$ module includes implementations in the sparse, constrained formulation. Table \ref{tab:random_walk_mixing_rate} provides a broad overview of their respective mixing rates and general description. 
\begin{table}[ht]
    \centering
    \begin{tabularx}{\linewidth}{>{\centering\arraybackslash}p{0.25\linewidth}>{\centering\arraybackslash}p{0.15\linewidth}>{\arraybackslash}p{0.5\linewidth}}
        \toprule
        \textbf{Random Walk} & \textbf{Mixing Rate} & \textbf{Computational Complexity} \\
        \midrule
        \addlinespace
         Ball Walk \cite{vempala2005} & $\tau(d^2 R^2/r^2)$ & \fontsize{9}{8}\selectfont The Ball Walk includes minimal matrix multiplication and gaussian vector calculations.\\
         \addlinespace
         Hit-and-Run \cite{lovasz1999} & $\tau(d^2 R^2/r^2)$ &  \fontsize{9}{8}\selectfont The Hit-and-Run computes binary search calculate a random chord in the polytope.\\
         \addlinespace
         Dikin Walk \cite{DBLP:journals/corr/SachdevaV15} & $\tau(nd)$ & \fontsize{9}{8}\selectfont The Dikin Walk uses sparse linear solvers and sparse Cholesky decomposition for computing determinant and inverse square root of Hessian matrix.\\
         \addlinespace
         Vaidya Walk \cite{JMLR:v19:18-158} & $\tau(n^{1/2}d^{3/2})$ & \fontsize{9}{8}\selectfont The Vaidya Walk builds upon Dikin Walk and uses sparse Cholesky decomposition and sparse inverse algorithm from \cite{DBLP:journals/corr/abs-2202-01908} to compute leverage scores.\\
         \addlinespace
         John Walk \cite{JMLR:v19:18-158} & $\tau(d^{2.5})$ & \fontsize{9}{8}\selectfont The John Walk builds upon the Vaidya Walk to solve an optimization task with fixed-point iteration, computing a weighted leverage score each iteration.\\
         \addlinespace
         Lee Sidford Walk \cite{DBLP:journals/corr/abs-1911-05656} & $\tau(d^{2})$ & \fontsize{9}{8}\selectfont The Lee Sidford Walk also solves an optimization task using gradient descent, computing a weighted leverage score each iteration. \\
        \bottomrule
    \end{tabularx}
    \caption{In each, $d$ refers to the dimension of the polytope and $n$ refers to the number of boundaries ($\mathcal{K}_1$ dimensions). In the first 2 walks, $R^2/r^2$ means where the convex body contains a ball of radius $r$ and is mostly contained in a ball of radius $R$. }
    \label{tab:random_walk_mixing_rate}
\end{table}
\subsection{Prepossessing Algorithms}
\package comes with 2 preprocessing algorithms: initialization and facial reduction.

\paragraph{Initialization} If the user cannot specify a point inside of the polytope to start, \package provides a class to compute an initial point well within the polytope for both the full-dimensional formulation and constrained formulation. In $\mathcal{K}_2$, this is akin to solving the linear program maximizing $\delta \in \mathbb{R}$ s.t. $Ax =b$ and $x \succeq_k \vec{\delta}$. In $\mathcal{K}_1$, this is equivalent to maximizing $\delta \in \mathbb{R}$ such that $Ax + \delta \cdot \mathbbm{1} \le b$ where $\mathbbm{1}$ is the all-ones vector $\in \mathbb{R}^n$.

\paragraph{Facial Reduction} We adopt the facial reduction algorithm implementation from \cite{drusvyatskiy2017many} and \cite{im2023revisiting}. The facial reduction algorithm deals with cases of degeneracy in the polytope. In the constrained formulation $\mathcal{K}_2 = \{x \in \mathbb{R}^d \ | \ Ax = b, x \succeq_k 0\}$, degeneracy occurs when there is a lack of strict feasibility in the polytope: there does not exist an $x \in \mathbb{R}^d$ such that $Ax = b$ and $x \succ_k 0$. The facial reduction algorithm eliminates variables in the last k dimensions fixed at $0$. It is far more commonly used in linear programming than MCMC sampling. Degeneracy exists in polytopes when the lower-dimensinoal polytope is embedded in a higher dimension. For example, if one attempts to use MCMC algorithms to sample from a 2-dimensional hypercube in 5 dimensions, the proposal distributions which generate an ellipsoid in 5 dimensions will certainly reject each proposal. It is crucial that one projects the polytope back into the 2 dimensional form to run the MCMC sampler efficiently. Facial Reduction accomplishes this task, ensuring numerical stability for sampling. 

\subsection{Package Comparison}
\begin{table}[ht]
    \centering
    \caption{\textbf{Comparison of \package 
    with other MCMC sampling libraries.}\\} 
    \begin{adjustbox}{width=0.8\textwidth}
    \begin{tabular}{@{}llllll@{}}
         \toprule
        & \package & \textsf{Volesti} \cite{Chalkis_2021}  & \textsf{WalkR}\cite{Yao2017} & \textsf{Polyrun} \cite{CIOMEK2021100659} \\
        \midrule
        Constrained Formulation & \greencheck & \redcross & \greencheck & \greencheck \\
        Sparse Friendly & \greencheck & \redcross & \redcross & \redcross \\
        C++ Implementation & \greencheck & \greencheck & \redcross & \redcross \\
        Facial Reduction Algorithm & \greencheck & \redcross & \redcross & \redcross \\
        Dikin Walk & \greencheck & \greencheck & \greencheck & \redcross \\
        Vaidya Walk & \greencheck & \greencheck & \redcross & \redcross \\
        John Walk & \greencheck & \greencheck & \redcross & \redcross \\
        Lee-Sidford Walk & \greencheck & \redcross & \redcross & \redcross \\
        \bottomrule
    \end{tabular}
    \end{adjustbox}
    \label{table:package_checkmarks}
\end{table}
Table~\ref{table:package_checkmarks} contrasts the features of \package with various other open-source sampling libraries. \package is the first open-source package that enables users to leverage sparsity and use constrained formulations of the polytopes to generate points uniformly with fast MCMC algorithms. We are also one of the first to primarily focus on barrier-based MCMC samplers. \package includes a C++ implementation with corresponding Python wrapper code. Conversely, \textsf{Volesti} is implemented in C++ with some of its code represented in the Python library \textsf{Dingo}. \textsf{Polyrun} only works on Java and \textsf{WalkR} on R. Notably, \textsf{WalkR} was removed from the CRAN repository, motivating further open source MCMC sampling development. 
\section{Experiments}
The technical experimental details are listed in the Appendix section. We test \package on 5 \netlib dataset polytopes and 3 structured families of polytopes: simplex, hypercube and Birkhoff polytope, mirroring experiments from \cite{DBLP:journals/corr/abs-2202-01908}. We test the per-iteration cost, mixing cost, and total sampling cost in each instance. We sample upwards of $10^5$ dimensions for the per iteration cost and $10^3$ for the mixing cost for the structured polytopes.

Our experiments reveal improved mixing costs for the barrier-based walks over the standard walks on the 3 structured polytopes. We demonstrate improvements in runtime against the \textsf{Volesti}  implementation in both formulations. Our experiments verify the optimization done to improve open-source MCMC sampling algorithms efficiency. We also perform a uniformity test on our package to show its stationary distribution is approximately correct.

\section{Conclusion}
We presented the features of \package, demonstrating its improvement in runtime and its application of novel methods for preprocessing polytopes and sampling in the sparse constrained form. Our package supports Python and C++ development. \package is design to improve existing MCMC algorithms with added rigor in initialization and flexibility for sparse polytope formulations. We plan to keep updating the library with further code optimizations and new MCMC methods as cutting edge research is produced. 


\section*{Acknowledgement}
Much of the work was done while Yuansi Chen was an assistant professor in the Department of Statistical Science at Duke University. Both authors are partially supported by NSF CAREER Award DMS-2237322, Sloan Research Fellowship and Ralph E. Powe Junior Faculty Enhancement Awards. 

\bibliography{main}

\newpage
\appendix
\section{MCMC Sampling Details}
\subsection{Full Dimensional Form Setup}
We examine the setup of general barrier walks in $\mathcal{K}_1 = \{v \in \mathbb{R}^{d-n} \ | \tilde{A}v \le \tilde{b}\}$ where $\tilde{A} \in \mathbb{R}^{k \times (d-n)}$ and $\tilde{b} \in \mathbb{R}^{k}$. Note that we chose these dimensions in terms of their higher dimensional $\mathcal{K}_2$ formulation dimensions. Define the weighted barrier function
\begin{align*}
    F(v) \defn -\sum_{i=1}^{k} w_i \log(b_i - a_i\tp v)
\end{align*}
Its Hessian is
\begin{align*}
    H(v) \defn \sum_{i = 1}^k w_i\frac{a_i a_i\tp}{(b_i - a_i\tp v)^2} = \tilde{A}\tp S_v^{-1}W_vS_v^{-1}\tilde{A}
\end{align*}
where $S_v = \text{diag}(\tilde{b} - \tilde{A}v)$ is the slack matrix and $W_v = \text{diag}(w) \in \mathbb{R}^{k \times k}$ is the weights. The different barrier walks mainly differ in the choice of $w$ which is computed at each step. The proposal distribution at $v$ is
\begin{align*}
    \mathcal{N}\left(v, \frac{r^2}{c^2} H(v)^{-1}\right)
\end{align*}
$r$ is a hyperparameter chosen to specify the relative size of the radius while $c$ is a walk-specific variance correction term, often in the function $n$, $d$, and $k$. Note that for $H(v)$ to be invertible, we need more constraints than the dimension, which will be the case if the interior is nonempty and we have a bounded polytope. Table \ref{tab:walk_weight} summarizes the formulation of $w \in \real^{k}$, where $W_v = \diag(w)$ for each of the barrier walks in the full dimensional form.
\begin{table}[ht]
    \centering
    \renewcommand{\arraystretch}{2} 
    \begin{tabularx}{1\linewidth}{|>{\centering\arraybackslash}m{0.2\linewidth}|>{\centering\arraybackslash}m{0.74\linewidth}|}
        \hline
        \textbf{Walk} & \textbf{Weight Formulation} \\
        \hline
        Dikin Walk & \small $ \mathbbm{1}$\\
        Vaidya Walk & \small $\sigma\left(S_v^{-1}\tilde{A}\right) + \frac{d-n}{k}$\\
        John Walk & \small $\underset{w \in \real^k}{\text{arg min}}\left\{ \sum_{i=1}^k w_i - \frac{1}{\alpha_\text{J}} \text{log det}(\tilde{A}\tp S_v^{-1} W_v^{\alpha_{\text{J}}} S_v^{-1} \tilde{A}) -\beta_{\text{J}}\sum_{i=1}^{k} \log w_i\right\}$\\
        Lee-Sidford Walk & \small $\underset{w \in \real^k, w \ge 0}{\text{arg min}}\left\{ \left(1 -\frac{2}{q}\right)\sum_{i=1}^k w_i - \text{log det}(\tilde{A}\tp S_v^{-1} W_v^{1 - \frac{2}{q}} S_v^{-1} \tilde{A}) \right\}$
        \\ & \\
        \hline
    \end{tabularx}
    \caption{Weights Formulation for Barrier Walks in Full Dimensional Form}
    \label{tab:walk_weight}
\end{table}

For the Dikin Walk, we do not include weights. For the Vaidya Walk, $\sigma(A) = \diag(A (A\tp A)^{-1} A\tp)$ or the leverage score. This computes the Vaidya volumetric barrier weights. Both the John Walk and Lee Sidford Walk find their weights by solving a convex optimization program. For the John Walk, $\beta_{\text{J}} = (d-n)/2k$ and $\alpha_{\text{J}} = 1-1/\text{log}_2(1/\beta_{\text{J}})$. For the Lee Sidford Walk, $q = 2(1 + \log(k))$. In \package, we develop equivalent formulations in $\mathcal{K}_2$ formulation, which are similar in derivation to $\mathcal{K}_1$.
\subsection{Relationship between \texorpdfstring{$\mathcal{K}_1$} -\ and \texorpdfstring{$\mathcal{K}_2$} -\ via QR Decomposition}
We would like to understand how $\mathcal{K}_1$ and $\mathcal{K}_2$ are related. We begin with $\mathcal{K}_2 = \{x \in \mathbb{R}^d \ | \ Ax = b, x \succeq_k 0\}$ where $A \in \mathbb{R}^{n \times d}$ and $b \in \mathbb{R}^n$. We assume $A$ is in general position. We have $n$ linear constraints in a $d$-dimensional space, so the dimension of the affine subspace $Ax = b$ is $d - n$. To find its corresponding full-dimensional form, we use QR decomposition $A\tp = QR$ where $Q \in \mathbb{R}^{d \times d}$ and $R \in \mathbb{R}^{d \times n}$. If we let $x = Qv$, $v \in \mathbb{R}^{d - n}$ then $\mathcal{K}_2$ is equal to
\begin{align*}
    R\tp Q\tp Qv &= b\\
    Qv &\succeq_k 0
\end{align*}
It follows from orthogonality that 
\begin{align*}
    R\tp v &= b\\
    Qv &\succeq_k 0
\end{align*}
We can write 
\begin{align*}
    Q = \bmat{Q_1 & Q_2}, R = \bmat{R_1 \\ 0_{n \times (d - n)}}, v = \bmat{v_1\\v_2}
\end{align*}
where $Q_1 \in \mathbb{R}^{d \times n}$, $Q_2 \in \mathbb{R}^{d \times (d - n)}$, $R_1 \in \mathbb{R}^{n \times n}$, $v_1 \in \mathbb{R}^n$, and $v_2 \in \mathbb{R}^{d-n}$. $R_1$ is invertible since $A$ is in general position. It follows from the first equality that 
\begin{align*}
    v_1 &= R_1^{-\top}b\\
    Q_1v_1 + Q_2v_2 &\succeq_k 0
\end{align*}
We arrive at the following full-dimensional formulation of the polytope ($\mathcal{K}_1$):
\begin{align*}
    \tilde{A}v_2 \le \tilde{b}
\end{align*}
where $\tilde{A} \in \mathbb{R}^{k \times (d - n)}$ is the last k rows of $-Q_2$ and $\tilde{b} \in \mathbb{R}^k$ is the last k rows of $Q_1R_1^{-\top}b$. For notational purposes, we denote $\bar{Q_2} = Q_2[d-k+1: , :]$ and $\bar{Q_1} = Q_1[d-k+1: , :]$, each denoting the last k rows of the matrix. Note that $k$ determines the number of inequality constraints and $d - n$ determines the dimension of the full-dimensional formulation. With a generalized barrier walk, $v \in \real^{d-n}$ in the full-dimensional formulation has the next proposal:
\begin{align*}
    \mathcal{N}\left(0, \frac{r^2}{c^2} H_v^{-1}\right)\\
    v_{\text{next}} = v_{\text{prev}} + \frac{r}{c}H_v^{-\frac{1}{2}} \xi
\end{align*}
where $\xi$ is a $d - n$ dimensional standard Gaussian vector and $H_v^{-1}$ is the second derivative of the generalized weighted barrier. 
\begin{align*}
    H_v &= \Bar{Q_2}\tp S_v^{-1} W_v S_v^{-1} \Bar{Q_2}
\end{align*}
$S_v \in \mathbb{R}^{k \times k}$ is the slack matrix with diagonal entries $b - \tilde{A}v = \text{diag}(\bar{Q_2}v + \bar{Q_1}R_1^{-\top}b)$. We note that $x$ and $v$ are related via $x = Q_2v + Q_1R_1^{-\top}b$, $S_v = \text{diag}(x[d-k:])$. Thus, the update in the original constrained formulation (or high dimensional form) is
\begin{align*}
    Q_2v_{\text{next}} &= Q_2v_{\text{prev}}+ \frac{r}{c} Q_2 H_v^{-\frac{1}{2}} \xi
\end{align*}
Therefore, the next point is sampled from a Gaussian  centered at $Q_2v$. with covariance 
\begin{align*}
    \left(\frac{r}{c}\right)^2Q_2 H_v^{-1}Q_2^T
\end{align*}
This linkage fundamentally connects the full-dimensional formulation and constrained formulation together with the QR decomposition of $A$. 
\subsection{Implementation Details in \texorpdfstring{$\mathcal{K}_2$} -\ Formulation}
In Algorithm \ref{algo:barrier_description}, we show the general formula for running the barrier walks in $\mathcal{K}_2$ form.\\  
\begin{algorithm}[ht]
  \KwIn{Parameter $r$ and $x_0 \in \mathcal{K}$}
  \KwOut{Sequence $x_1, x_2, \ldots$}
  \BlankLine
  \For{$i=0, 1, \ldots $}{
     {%
      \quad \textbf{Proposal step}:
      Draw $z_{i+1} \sim
      \mathcal{N}\left(x_{i}, \frac{r^2}{c^2}M_{x_{i}}^{\dagger}\right)$ \\
      \quad \textbf{Accept-reject step}:\\
      \quad \quad \emph{if} $z_{i+1} \notin \mathcal{K}$
      \emph{then} {$ x_{i+1} \gets
        x_i $ } \quad \% {\footnotesize{\emph{reject an infeasible
      proposal}}} \\
      \quad \quad \emph{else}\\
      \quad\quad\quad compute
       $\alpha_{i+1} = \displaystyle\min \braces{1,
          {p_{z_{i+1}}(x_{i+1})}/{p_{x_{i+1}}(z_{i+1})}
        }$\\
        \quad \quad \quad With probability $\alpha_{i+1}$ accept the proposal:
        $x_{i+1} \gets
        z_{i+1} $\\
        \quad \quad \quad With probability $1-\alpha_{i+1}$ reject the proposal:
        $x_{i+1} \gets
        x_i$}
    }
  \caption{Barrier Walk with parameter $r$}
  \label{algo:barrier_description}
\end{algorithm}

In the algorithm, $c$ represents a specific constant for each random walk and $M^\dagger_{x_i}$ is the covariance matrix that represents the pseudo inverse of the Hessian of the specified barrier at $x_i$. We draw from a normal distribution in our proposal step and use a Metropolis-Hastings accept reject step with a normal density to ensure a uniform stationary distribution. To find an equivalent algorithm between the $\mathcal{K}_1$ and $\mathcal{K}_2$ formulations, we must understand how to modify the proposal step and the accept-reject step to account for the constrained formulation of the polytopes.
\subsubsection{Proposal Step}
We define $g(x) = S_x^{-1}W_xS_x^{-1}$ as the local metric. $S_x^{-1} \in \real^{d \times d}$ is the inverse slack matrix in higher dimensional form where $S_x^{-1} = \diag(0,... ,1/x_{d-k+1}, ..., 1/x_{d})$. $W_x \in \real^{d \times d}$ is the high dimensional weight determined by the walk choice (Dikin, Vaidya, etc) which is a matrix with block $W_v$ in the bottom right corner. When we formulate $M$, we also note that $\bar{Q_2}\tp S_v^{-1}W_vS_v^{-1}\bar{Q_2}$ (the lower-dimensional form) is identical to $Q_2\tp g(x) Q_2$ where the first $n - k$ diagonal elements of $g(x)$ is 0 and the last $k$ elements are the diagonal entries of $S_v^{-1}W_vS_v^{-1}$.  In implementation, we add $\epsilon$ to the diagonal entries of $g(x)$ for invertibility.

In the first step, we generate the next point from a proposal distribution 
\begin{align*}
    \mathcal{N}\left(0, \frac{r^2}{c^2} M^\dagger \right)\\
    x_{\text{next}} = x + \frac{r}{c} \sqrt{M^\dagger}\zeta
\end{align*}
where $\zeta$ is a standard normal vector $\in \real^{d}$. The covariance from proposal distribution is derived from the local metric $g(x)$. Equivalently, $M$ is the orthogonal projection of $g(x)$ onto the nullspace of $A$. The paper from \cite{DBLP:journals/corr/abs-2202-01908} shows an identical formula:
\begin{align}
    M^\dagger &= g(x)^{-1} - g(x)^{-1}A\tp(A g(x)^{-1} A\tp)^{-1}Ag(x)^{-1}
    \label{eq:kook_inverse1}
\end{align}
Equation \ref{eq:kook_inverse1} has the added benefit that one does need not to do QR decomposition on A. However, we originally specified $g(x)$ to include zero diagonal elements. In our formulation, we add $\epsilon$ error to the diagonals to ensure that the formulation is still viable.  Moreover, it can be shown that the square root of $M^\dagger$ is just $g(x)^{-1/2} - g(x)^{-1}A\tp (Ag(x)A\tp)^{-1} A g(x)^{-1/2}$. Previously, we have proven that both proposal distributions in $\mathcal{K}_1$ and $\mathcal{K}_2$ forms are identical.
\subsubsection{Accept-Reject Step}
Next, we must apply a Metropolis-Hastings accept reject step. Let $z$ be the proposed next step and $x$ is the current iterate. We set $\alpha = \min(1, p_z(x)/p_x(z))$. $p_z(x)$ refers to the probability density of the proposal distribution centered at z to the point x. We set the next iterate to $z$ with probability $\alpha$ and reject the proposal with probability $1 - \alpha$. Note that the probability density function is from a multivariate normal distribution where
\begin{align*}
    p_x(z) &= \sqrt{\text{pdet}(M_x)}\exp\left(-\frac{c^2}{r^2}(z-x)\tp M_x(z-x)\right)
\end{align*}
In the full dimensional $\mathcal{K}_1$ form, we show that from $v$ to $w$, it is 
\begin{align*}
    p_v(w) &= \sqrt{\text{det}(H_v)}\exp\left(-\frac{c^2}{r^2}(w-v)\tp H_v (w-v)\right)
\end{align*}
The paper~\cite{DBLP:journals/corr/abs-2202-01908} proves the following equation for $\text{pdet}(M)$:
\begin{align}
    \text{pdet}(M) &= \frac{\det(g(x))\det(Ag(x)^{-1}A\tp)}{\det(AA\tp)}
    \label{eq:kook_pdet}
\end{align}
This is where $\text{pdet}$ refers to the pseudo determinant. By showing equivalent procedures for both the proposal distribution accept-reject step for a generalized barrier walk, we complete the description for the implementation for the $\mathcal{K}_2$ sparse constrained formulation. 
\section{Facial Reduction Details}
This facial
reduction algorithm description follows from \cite{drusvyatskiy2017many} and \cite{im2023revisiting}. Given $A \in \real^{n \times d}, b \in \real^n$, we are interested in the feasible region $\cF$
\begin{align}
  \label{eq:standard_form2_feasible_region}
  \cF \defn \braces{x \in \real^d : A x = b, x \succeq_{\cC} 0}.
\end{align}
where $\cC = \braces{x \in \real^d, x_{d-k+1} \geq 0, \ldots, x_d \geq 0}$ and the notation $x \succeq_{\cC} 0$ means that $x_{d-k+1} \geq 0, \ldots, x_n \geq 0$.
We then show Theorem 3.1.3 in~\cite{drusvyatskiy2017many}.
\paragraph{face.} Let $K \subset \real^n$ be a convex set. A convex set $F \subseteq K$ is called a face of $K$, if for all $y, z \in K$ with $x = \frac{1}{2} (y+z) \in F$, then $y, z \in F$. 
\paragraph{minimal face.} Given a convex set $C \subseteq K$, the minimal face for $C$ is the intersection of all faces containing the set $C$. 
\paragraph{facial reduction.} The facial reduction is a process of identifying the minimal face of $\real^n_+$ (more generally can be a cone) containing the feasible set $\cF$. By finding the minimal face, we can work with a problem that lies in a smaller dimensional space and that satisfies strict feasibility.
\paragraph{Theorem of the alternative} For~\eqref{eq:standard_form2_feasible_region}, exactly one of the following statement holds
\begin{enumerate}
  \item $\cF$ is strictly feasible. i.e., there exists $x \in \real^d$ such that $x_{d-k + 1} > 0, \ldots, x_d > 0$.
  \item there exists $y \in \real^n$ such that 
  \begin{align*}
    0 \neq z \defn A\tp y \succeq_{C^*} 0 \text{ and } b\tp y = 0
  \end{align*}
  where $C^* \defn \braces{x \in \real^d, x_1=0, \ldots, x_{d-k}=0, x_{d-k+1} \geq 0, \ldots, x_d \geq 0}$. 
\end{enumerate} 

\paragraph{Summary of the Facial Reduction Algorithm } 
\begin{enumerate}
  \item (FindZ) According to the theorem of the alternative, we can find $z, y$ such that
  \begin{align*}
    0 \neq z \defn A\tp y \succeq_{C^*} 0 \text{ and } b\tp y = 0.
  \end{align*}
  \item (PickV) Since $x\tp z = (Ax)\tp y = b\tp y = 0$ and each term is nonnegative, we have 
  \begin{align*}
    x_i z_i = 0, \forall i.
  \end{align*}
  Hence if $z_i > 0$, then it has to be that $x_i = 0$. Let $s_z = \abss{\braces{i : z_i > 0}}$ be the cardinality of its support. Then we can write
  \begin{align*}
    x = \sum_{j=1}^{d-k} x_j e_j + \sum_{j = d-k+1}^d \mathbf{1}_{z_j = 0} x_j e_j,
  \end{align*} 
  where for $j \in [d-k+1, d]$, $x_j \geq 0$. Define the matrix with unit vectors for columns
  \begin{align*}
    V = \bmat{e_1 & e_2 & \cdots & e_{d-k} & e_{s_1} &\cdots & e_{s_{d-k-s_z}} } \in \real^{d \times (d - s_z)}.
  \end{align*}
We have the equivalent problem formulation using 
$$v \in C^{d - s_z} \defn \braces{v \in \real^{d-s_z}, x_{d-k+1} \geq 0, \ldots, x_{d-s_z} \geq 0 }$$
  \begin{align*}
    \cF = \braces{x \in \real^d : A x = b, x \succeq_{\cC} 0} = \braces{x = Vv \in \real^d, AV v = b, v \succeq_{C^{d-s_z}} 0} .
  \end{align*}
  \item (PickP) It is known that every facial reduction step results in at least one constraint being redundant. Let $P_{\bar{n}}: \real^n \to \real^{\bar{n}}, \bar{n} = \text{rank}(AV)$, be the simple projection that chooses the linearly independent rows of $AV$. Then we have the equivalent full problem  
  \begin{align*}
    \cF &= \braces{x \in \real^d : A x = b, x \succeq_{\cC} 0} \\
    &= \braces{x = Vv \in \real^d, AV v = b, v \succeq_{C^{d-s_z}} 0} \\
    &= \braces{x = Vv \in \real^d, P_{\bar{n}} A V v = P_{\bar{n}} b, v \succeq_{C^{d-s_z}} 0 }
  \end{align*}
\end{enumerate}
At the end, our feasible convex set $\cF$ is now in a smaller dimension, enabling us to sample with greater numerical stability. At this point, we may do QR decomposition to convert the sparse polytope into full-dimensional form or we may continue sampling in this form. 

\section{Experimental Details}
We test \package on structured families of polytopes and \netlib. We provide definitions for each of the polytopes that we sample from, mirroring the experiments from \cite{DBLP:journals/corr/abs-2202-01908}.\\

\textbf{Simplex}: The simplex is $\{x \in \mathbb{R}^d : \sum_{i=1}^d x_i = 1, x \ge 0\}$. If the constrained form is in $\real^d$, then the full-dimensional form is $\real^{d-1}$.\\

\textbf{Hypercube}: The hypercube is $\{x \in \mathbb{R}^{d} | -1 \le x_i \le 1 \text{ for all } i \in [d]\}$. If the constrained form is in $\mathbb{R}^d$, then its full-dimensional form is in $\mathbb{R}^{d/3}$.\\

\textbf{Birkhoff}: The Birkhoff polytope is the convex hull of all permutation matrices defined from \cite{DBLP:journals/corr/abs-2202-01908} as 
\begin{align*}
    B_d &= \{(\left(x_{ij}\right)_{(i, j) \in [d]} : \sum_{j} x_{ij} = 1 \text{ for all } i \in [d], \sum_{i} x_{ij} = 1 \text{ for all } j \in [d], x_{ij} \ge 0\}
\end{align*}
If the constrained form is in $\real^{d^2}$, its full-dimensional form is $\real^{(d-1)^2}$.\\

We test 5 random walk algorithms from \package on the 3 structured families of polytopes. We exclude the Lee Sidford Walk from our experiments due to the slow convergence of the gradient descent method, unlike the fixed point iteration method for finding the John Weights. We examine both the sparse, constrained formulation on the polytope and the dense, full-dimensional formulation, computed using QR decomposition. Finally, we compare our dense formulation with the \textsf{Volesti} package both in a per-iteration cost (or the time it takes for the MCMC sampler to move one step) and the mixing cost (number of steps before an independent sample) across the dimension.\\

\begin{figure}[ht]
    \centering
    \advance\leftskip-2cm
    \includegraphics[width = 9cm]{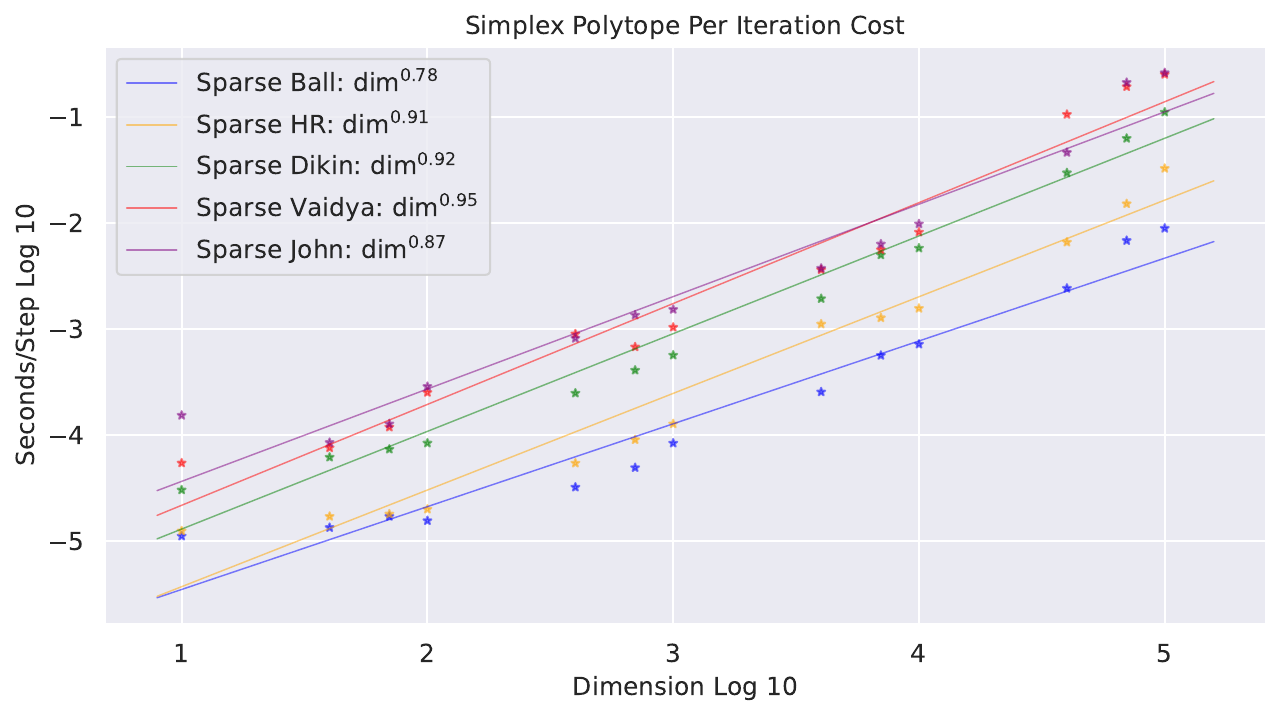}
    \includegraphics[width = 9cm]{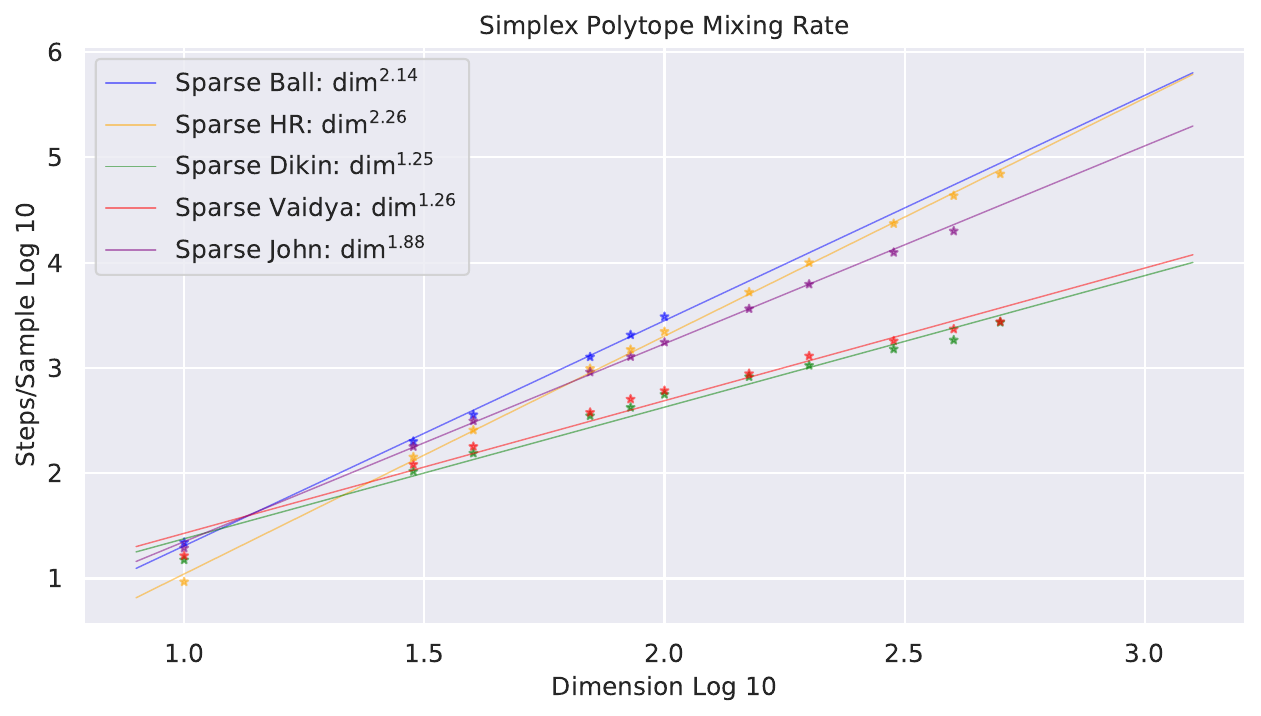}\advance\rightskip-2cm\\
    \advance\leftskip-2cm
    \includegraphics[width = 9cm]{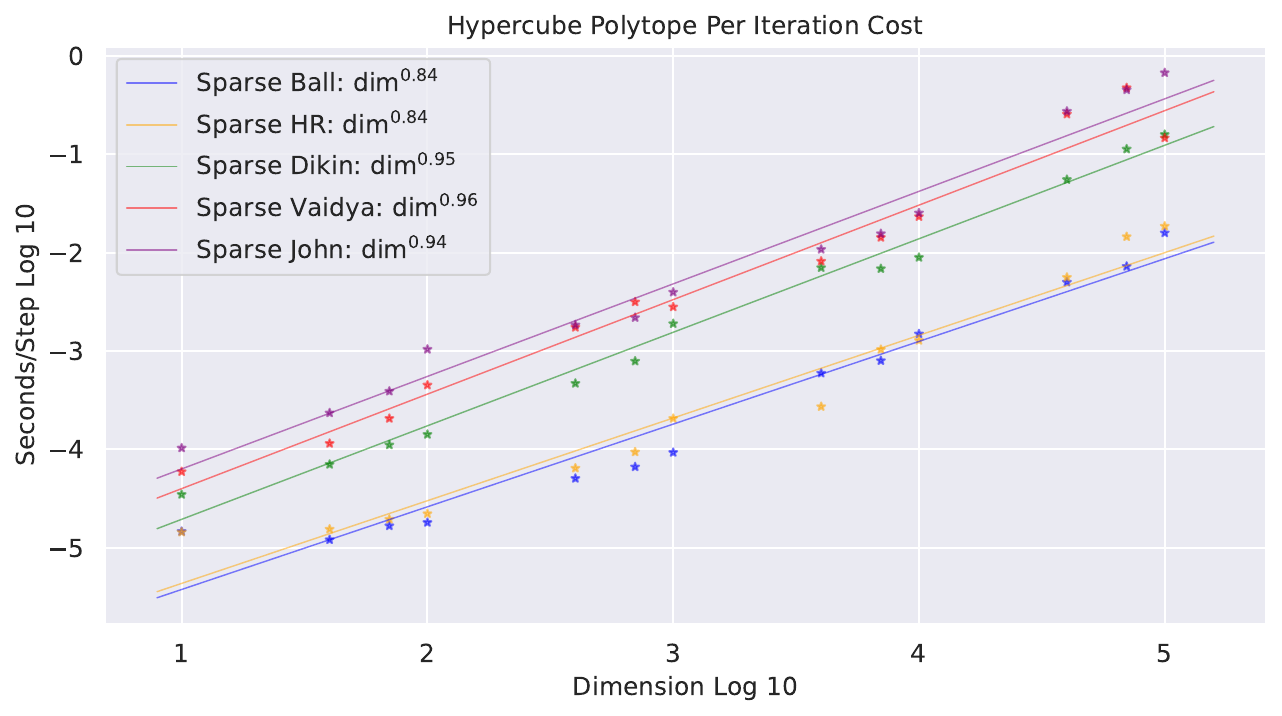}
    \includegraphics[width = 9cm]{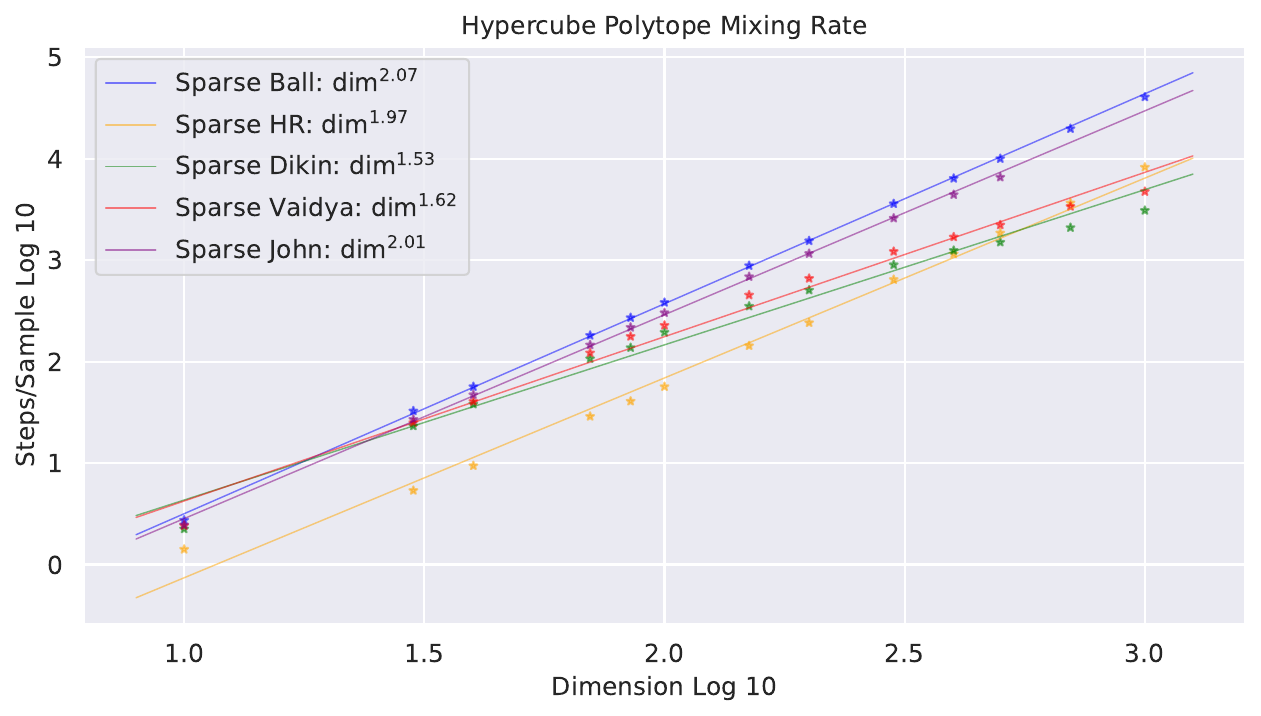}\advance\rightskip-2cm\\
    \advance\leftskip-2cm
    \includegraphics[width = 9cm]{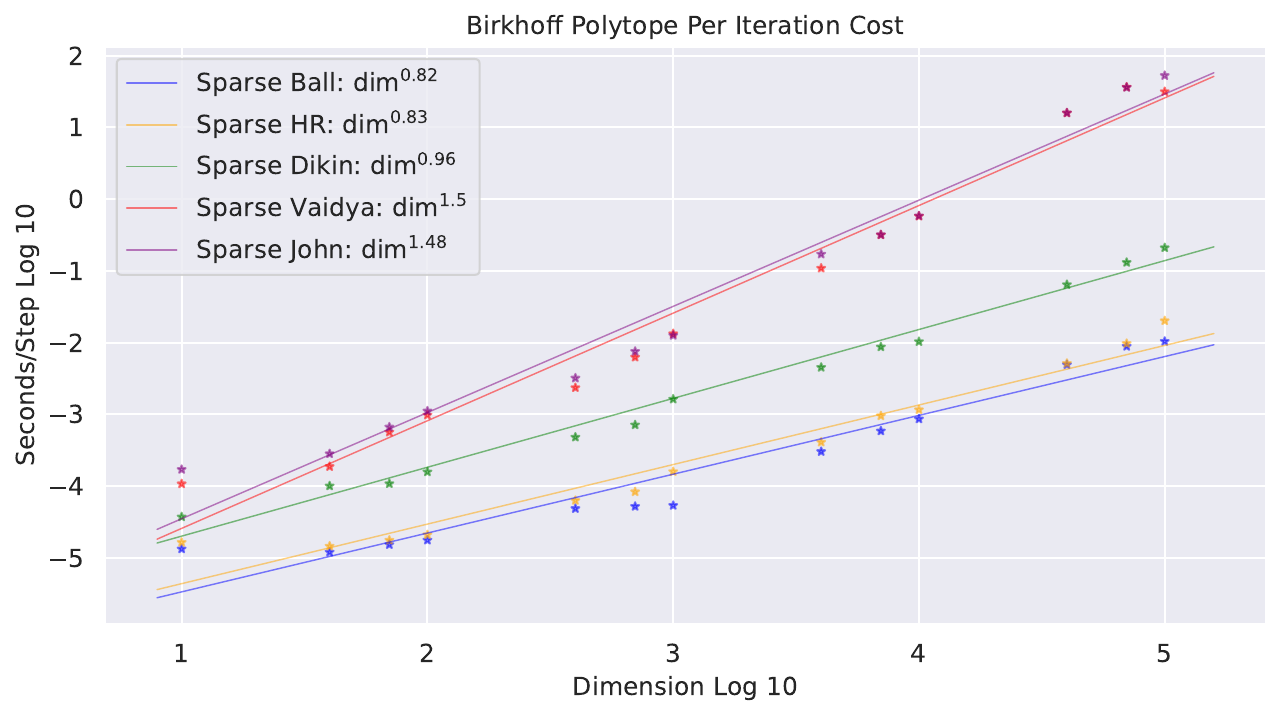}
    \includegraphics[width = 9cm]{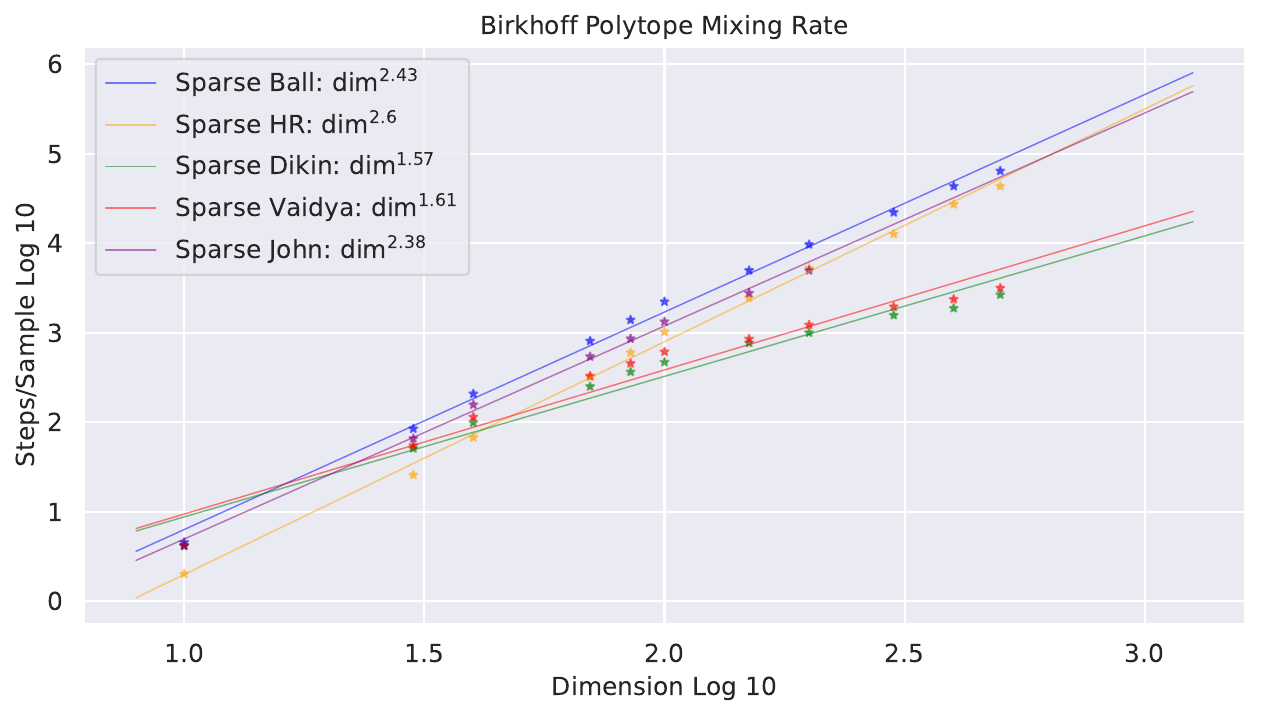}\advance\rightskip-2cm\\
    \caption{The per iteration / mixing rate cost for sparse implementation in \package}
    \label{table:sparse_res}
\end{figure}
For the per-iteration cost, we measure the time required for each MCMC sampler to move one step in the space across dimension, from $10^1$ to $10^5$ dimensions. We take 10 trials each of 500 steps to calculate the per iteration cost. For the mixing rate cost, we sample from the MCMC sampler until we have 500 independent samples, calculated using an effective sample size computation. We test the mixing rate from $10^1$ to $10^3$ dimensions. For our mixing rate, we sample $10 (d-n)^2$ steps, thinning our samples to only include every 10th. Since keeping each point takes a significant amount of storage, we measure the ESS of each trial and add everything together until we have 500 effective sample size. The barrier walks are approximately quadratic in mixing rate, so we sample enough steps in this process.\\

\begin{figure}[ht]
    \centering
    \advance\leftskip-2cm
    \includegraphics[width = 9cm]{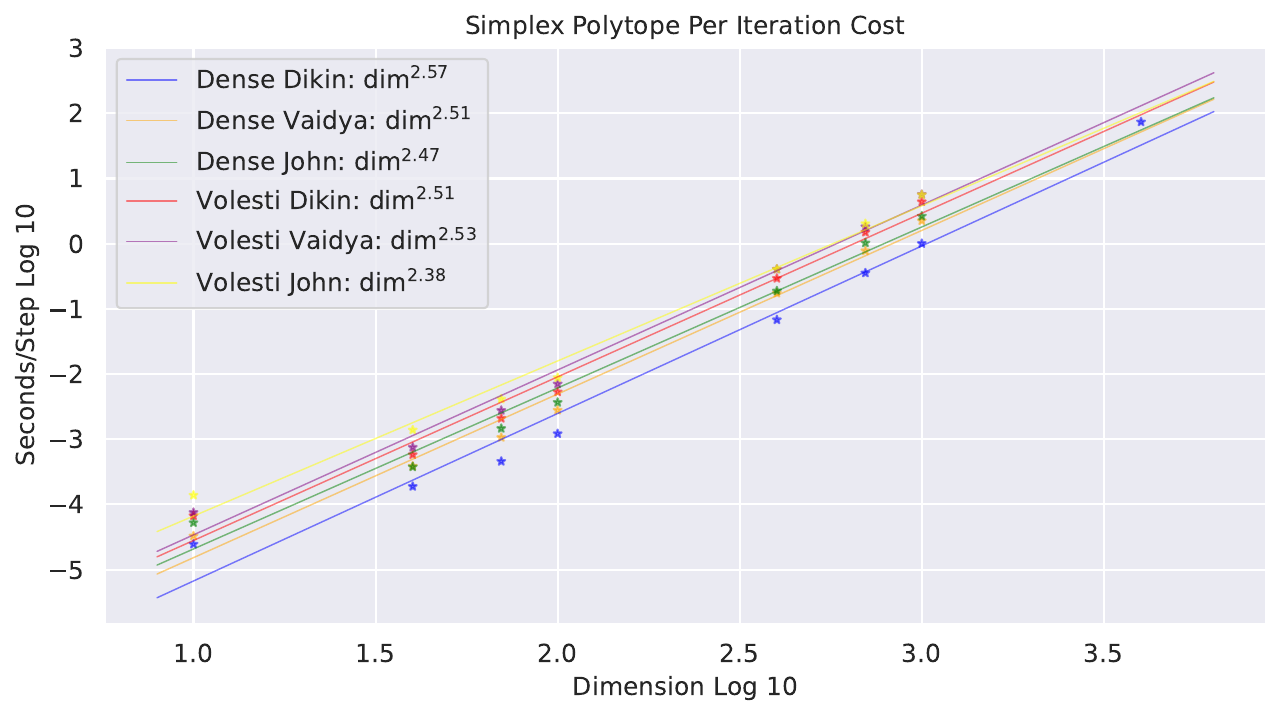}
    \includegraphics[width = 9cm]{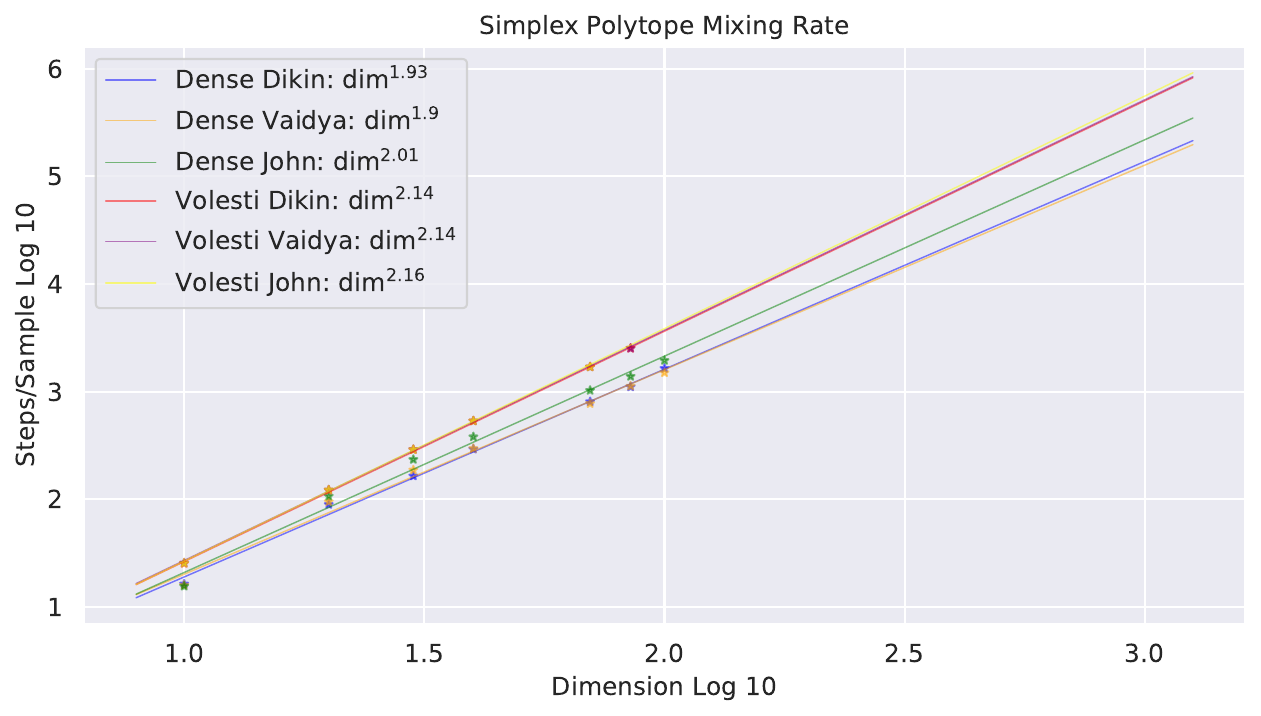}\advance\rightskip-2cm\\
    \advance\leftskip-2cm
    \includegraphics[width = 9cm]{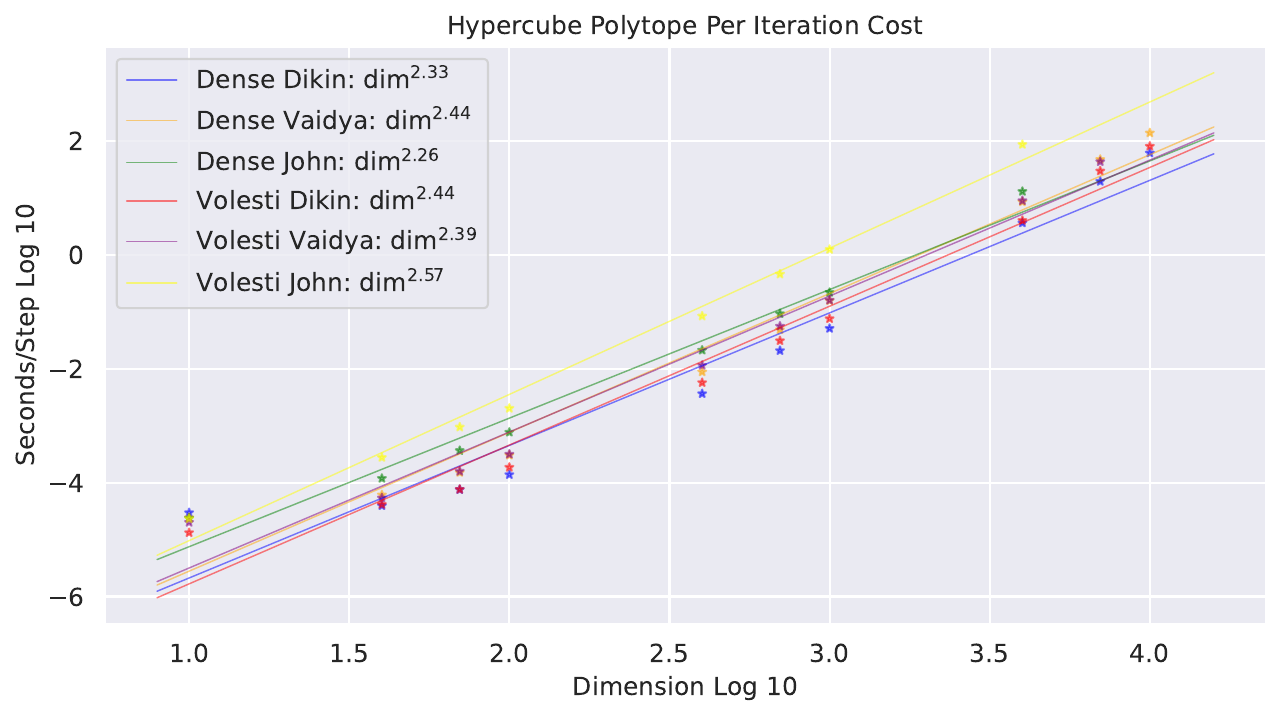}
    \includegraphics[width = 9cm]{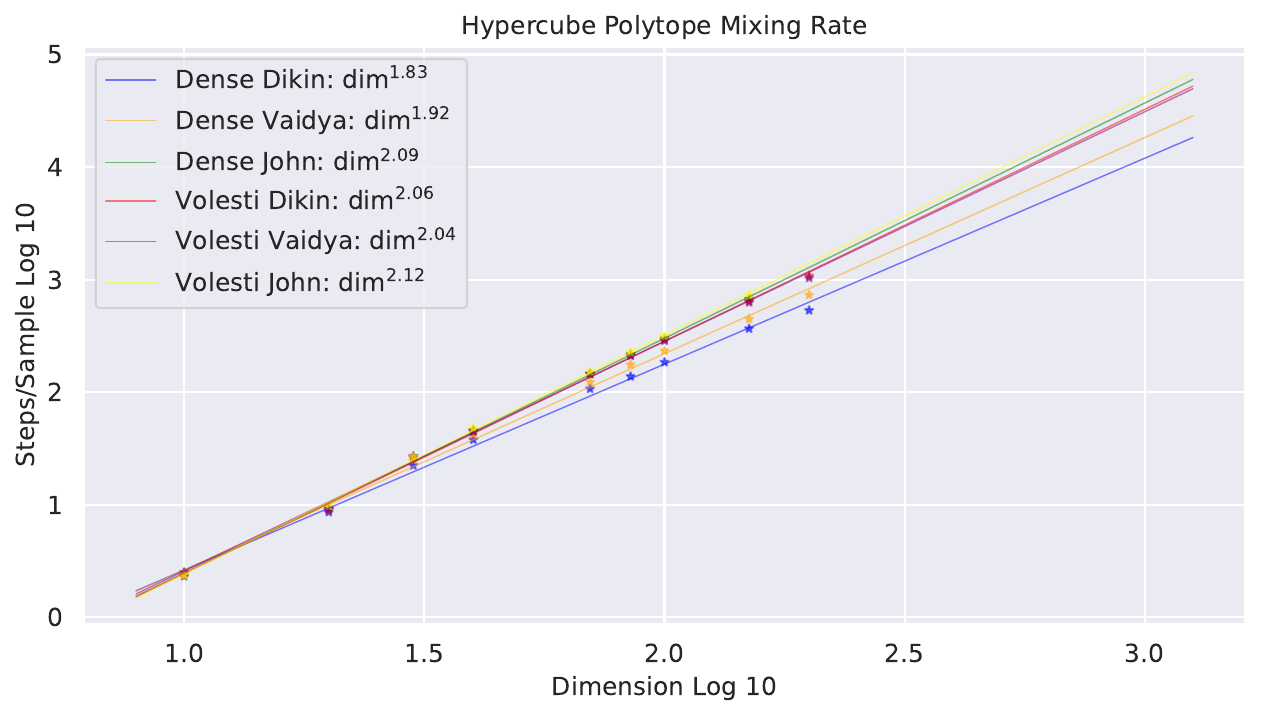}\advance\rightskip-2cm\\
    \advance\leftskip-2cm
    \includegraphics[width = 9cm]{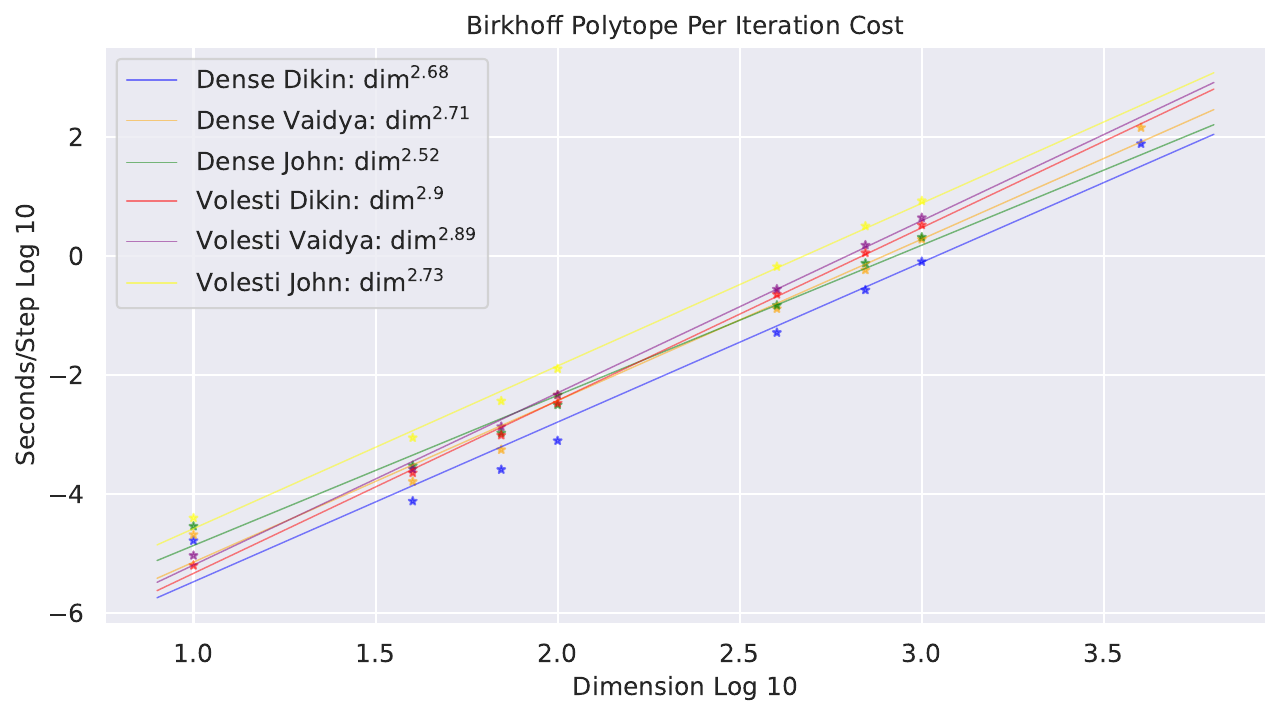}
    \includegraphics[width = 9cm]{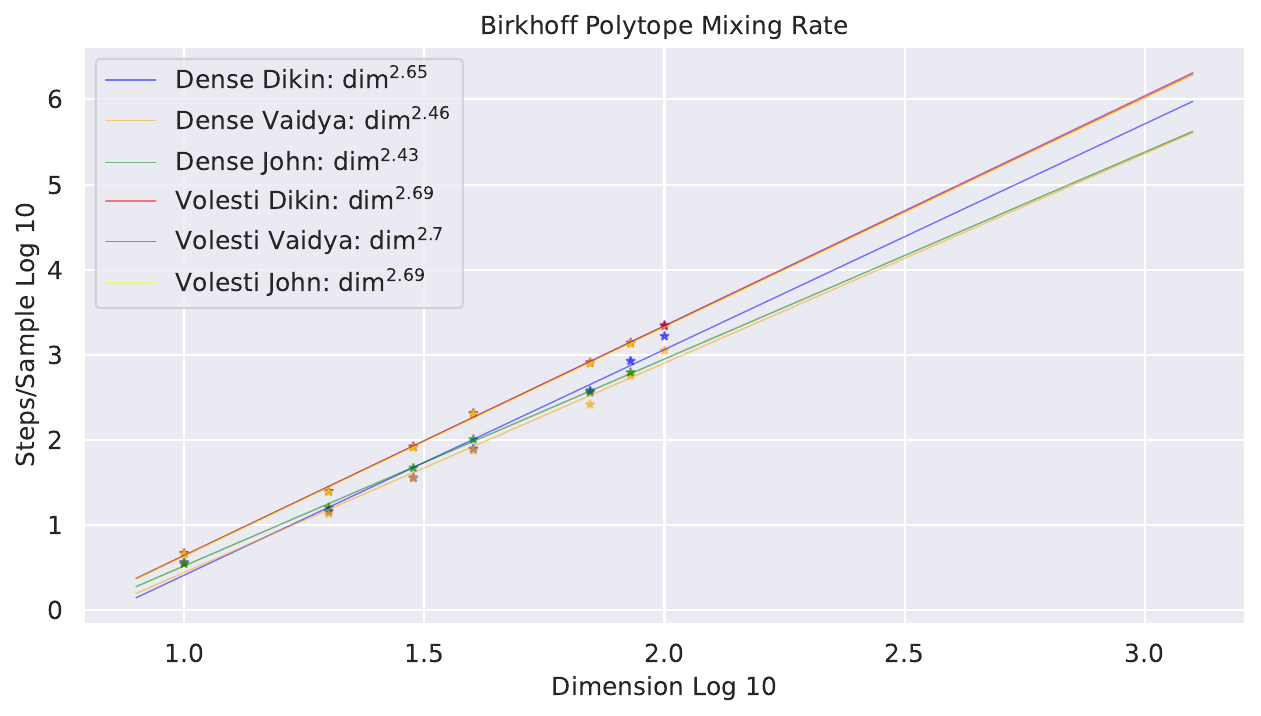}\advance\rightskip-2cm\\
    \caption{The per iteration / mixing rate cost for dense implementation in \package and \textsf{Volesti}}
    \label{table:dense_res}
\end{figure}

For both tests, we set a 24-hour limit with 16 GB of storage for each walk and dimension combination. The test will run until either the memory limit is exceeded, the time limit is exceeded, or it has completed the tasks. Table \ref{table:sparse_res} summarizes the sparse implementation results for the simplex, hypercube, and Birkhoff polytope.\\

The Ball Walk and the Hit-and-Run algorithm are demonstrably the fastest in each of the experiments on a per-iteration cost. However, their mixing rates show limitations, demonstrating the improvement that barrier walks have as dimensions grow. Notably, the Dikin and Vaidya Walk grow substantially slower, requiring less steps in the MCMC sampler to generate an independent random sample. The John Walk is slower in mixing time due to the larger constant in the mixing time operation. The sparse algorithms demonstrate vast improvements in per iteration cost. The sparse formulation demonstrates the ability of \package to sample from much higher dimensions than previous open source packages.\\

In Table \ref{table:dense_res}, we compare the \package against the \textsf{Volesti} package in the $\mathcal{K}_1$ formulation. While both are notably slower than the sparse implementation, \package has a clear advantage over \textsf{Volesti} for each of the 3 cases. Finally, the mixing rate costs were similar in performance between the two packages.\\

After testing on the structured families of polytopes, we apply our algorithms to the \netlib linear programming polytopes. Each are of the form $\mathcal{K} = \{x \in \mathbb{R}^d  |  Ax = b, x \ge 0\}$. To account for possible redundancies, we run the facial reduction algorithm on the expanded output to simplify the polytope and reduce degeneracy. Table \ref{tab:polytope_properties} reveals their characteristics.\\

\begin{table}[ht]
    \centering
    \begin{tabularx}{\linewidth}{|>{\centering\arraybackslash}X|>{\centering\arraybackslash}X|>{\centering\arraybackslash}X|>{\centering\arraybackslash}X|}
        \hline
        \textbf{Polytope} & \textbf{Constraints} & \textbf{Dimension} & \textbf{NNZ} \\
        \hline
        Adlittle & 56 & 138 & 424 \\
        Lotfi & 153 & 366 & 1136 \\
        Sctap1 & 300 & 660 & 1872 \\
        Scagr25 & 471 & 671 & 1725 \\
        Qap8 & 912 & 1632 & 7296 \\
        \hline
    \end{tabularx}
    \caption{Polytope Properties}
    \label{tab:polytope_properties}
\end{table}
Table \ref{tab:iteration_rate_netlib} reveals the per iteration cost for \package and \textsf{Volesti}.
\begin{table}[ht]
\centering
\begin{tabular}{|c|c|c|c|c|c|}
\hline
\textbf{Walk} & \textbf{Adlittle} & \textbf{Lotfi} & \textbf{Qap8} & \textbf{Scagr25} & \textbf{Sctap1} \\
\hline
Dense Dikin & 2.29 & 33.39 & 374.70 & 16.65 & 142.86\\
Dense Vaidya & 4.96 & 75.24 & 1124.53 & 87.84 & 333.46\\
Dense John & 6.67 & 82.92 & 1265.26 & 119.19 & 246.85\\
\hline
Sparse Ball & 0.02 & 0.06 & 0.53 & 0.09 & 0.05\\
Sparse Hit-and-Run & 0.03 & 0.08 & 0.63 & 0.16 & 0.07\\
Sparse Dikin & 0.58 & 2.07 & 14.51 & 1.37 & 3.28\\
Sparse Vaidya & 3.65 & 30.52 & $-$ & 7.31 & 24.59\\
Sparse John & 4.44 & 41.46 & $-$ & 8.61 & 32.54\\
\hline
Volesti Dikin & 7.32 & 460.37 & 802.35 & 496.88 & 161.03\\
Volesti Vaidya & 10.40 & 273.00 & 1033.00 & 234.54 & 358.12\\
Volesti John & 	36.45 & 272.66 & 1383.72 & 391.49 & 2645.5\\
\hline
\end{tabular}
\caption{Per Iteration Cost on \netlib Polytopes (Milliseconds / Step)} 
\label{tab:iteration_rate_netlib}
\end{table}
To compare \package and $\textsf{Volesti}$, we let each of the algorithms run until the 24 hour time limit, the 16GB memory limit, or if they reach 5000 steps in each test. Similar to prior results, the dense formulations of $\package$ outperform \textsf{Volesti}. Moreover, the sparse formulations of the Ball Walk, Hit-and-Run, and the Dikin Walk are the fastest. Sometimes, the sparse formulations are slower than their dense formulations. This is likely due to the very dense sparsity pattern of $A A\tp$ which induces additional computational efforts. A more concerning issue with \textsf{Volesti} was the initialization algorithm. In QAP8, Scagr25, and Lotfi, the radius calculation for the initial starting point encountered divide-by-zero errors, preventing the package from sampling multiple points from the polytopes. While \package was able of sampling from more narrow spaces in higher dimensions, \textsf{Volesti} was incapable. This motivates the need for a comprehensive initialization algorithm of \package over \textsf{Volesti}. \\

Finally, we conducted a uniformity test on the sparse implementation of the barrier based walks to check if the samples from \package truly represent a uniform sample of the polytope. We borrow the test procedure developed by \cite{DBLP:journals/corr/abs-2202-01908}. We plot the empirical cumulative distribution of the radial distribution to the power of $1/d$ with as the listed effective sample size above.
\begin{figure}[ht]
    \centering
    \includegraphics[width = 7.5cm]{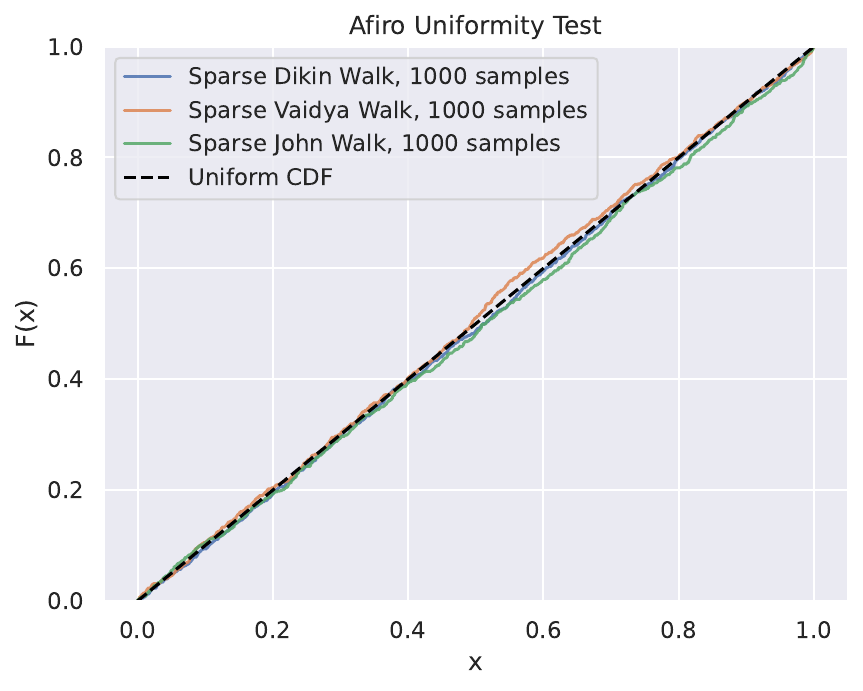}
    \includegraphics[width = 7.5cm]{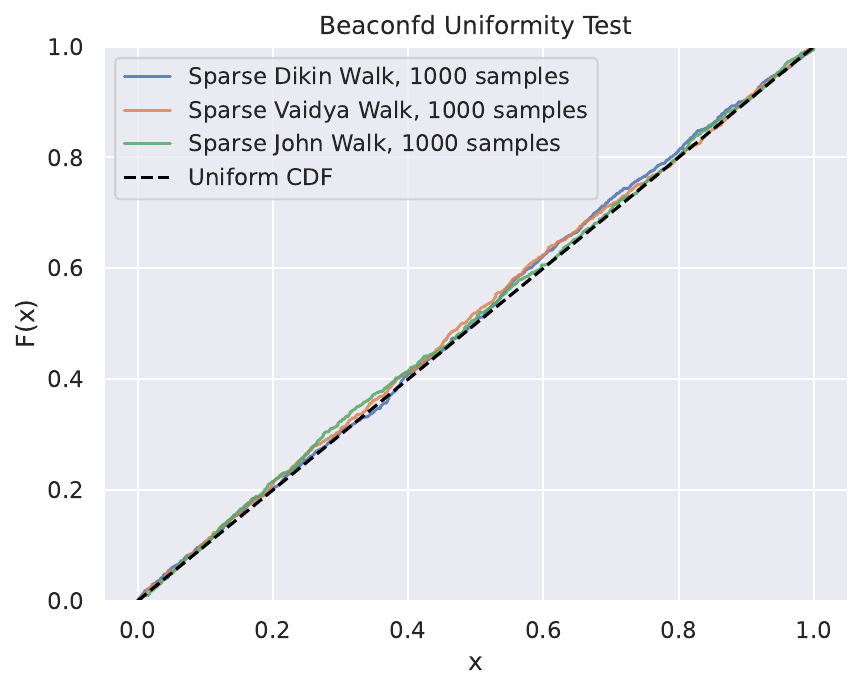}
    \caption{Uniformity Test on Afiro and Beaconfd \netlib Polytopes}
    \label{fig:uniformity_test}
\end{figure}
To generate a single independent sample, we run for $(d - n)^2$ steps each time. As explained in their paper, the empirical CDFs of the radial distribution to the $1/d$ power should match the CDFs of the uniform distribution over the polytopes. Figure \ref{fig:uniformity_test} demonstrates that $\package$ generates a stable uniform sample across the 2 \netlib polytopes: Afiro ($53 \times 77)$ and Beaconfd ($66 \times 99)$.  

\end{document}